\documentclass[12pt]{article}
\usepackage{cite}
\usepackage[pdftex]{graphicx}
\usepackage[cmex10]{amsmath}
\usepackage{url}
\usepackage{color}
\usepackage{listings}

\begin{document}
%
\title{A multi-terabyte relational database for \\ geo-tagged social network data}

\author{L\'aszl\'o Dobos, J\'anos Sz\"ule, Tam\'as Bodn\'ar, Tam\'as Hanyecz, \\
Tam\'as Seb\H{o}k, D\'aniel Kondor, Zs\'ofia Kallus, J\'ozsef St\'eger, \\
Istv\'an Csabai and G\'abor Vattay \\
Department of Physics of Complex Systems,  \\
E\"otv\"os Lor\'and University\\
Pf. 32, H-1518 Budapest, Hungary, \\
e-mail: dobos@complex.elte.hu}

\maketitle

\begin{abstract}
Despite their relatively low sampling factor, the freely available, randomly sampled status streams of Twitter are very useful sources of geographically embedded social network data. To statistically analyze the information Twitter provides via these streams, we have collected a year's worth of data and built a multi-terabyte relational database from it. The database is designed for fast data loading and to support a wide range of studies focusing on the statistics and geographic features of social networks, as well as on the linguistic analysis of tweets. In this paper we present the method of data collection, the database design, the data loading procedure and special treatment of geo-tagged and multi-lingual data. We also provide some SQL recipes for computing network statistics.
\end{abstract}
 


\section{Introduction}
Twitter is a micro-blogging web site that allows users to broadcast short textual status messages, \textit{tweets}, up to 140 characters which are automatically distributed to the \textit{followers} of the \textit{tweeting} user. Twitter also provides data streams for almost real-time data download at different random sampling rates. A stream with about 0.5\% sampling factor, the so called \textit{sprinkler}, can be accessed and downloaded for free and contains a representative subset of all tweets. Data is provided in the form of a continuous stream of JSON documents which is not the appropriate data format for large statistical analysis.

Twitter data is widely used in network sciences \cite{DBLP:dblp_conf/www/KwakLPM10} and previous work usually focused on results of the data analysis, and real-time stream processing \cite{DBLP:dblp_journals/firstmonday/BrunsL12, DBLP:dblp_conf/chi/HuLWWSM12, DBLP:dblp_journals/ijwbc/BenhardusK13} rather than how to collect and efficiently analyze large sets of archived data. A handful of papers discussed the possibility of building relational databases from social network data \cite{DBLP:dblp_conf/cn/WycislikW12}, while a few others even suggested special query languages \cite{DBLP:dblp_journals/debu/SmithB08, DBLP:dblp_conf/cidr/CohenEK13} for networks. A recent paper \cite{DBLP:dblp_journals/kbs/OussalahBCS13} described an architecture very similar to the system presented in this paper but built on a different platform.

Based on our earlier experiences with astronomical \cite{DBLP:conf/ssdbm/DobosBLSC12, DBLP:journals/corr/abs-1209-6490, DBLP:conf/edbt/DobosSBBCTMTJ11} and network measurement \cite{DBLP:conf/minenet/MatrayCHSDV07} data, we have built a data warehouse for geo-tagged Twitter data using Microsoft~SQL~Server~2012, a relational database engine. The database is designed to support a large variety of ad-hoc queries, and indices are defined especially for graph traversal, geographic queries and free text searches, as well as for large-scale statistics. In this paper we put emphasis on the description of the data loading process since regular updates to a multi-terabyte database by merging in new data is not a trivial task. For source code an more details on our project, please visit \url{http://www.vo.elte.hu/twitterdb}.

The structure of the paper is as follows. In Sec.~\ref{sec:data} the contents of the Twitter data stream and the caveats of data collection are explained. In Sec.~\ref{sec:database} we describe the database, the data loading procedure, as well as the ordinary and textual indexing of the data. Spatial indexing techniques are dissected in Sec.~\ref{sec:spatial}, whereas data access methods and visualization is touched in Sec.~\ref{sec:access}. We show a few sample queries to demonstrate the simplicity and efficiency of using the data warehouse in Sec.~\ref{sec:queries} and conclude the paper in Sec.~\ref{sec:conclusions}.

\section{Twitter data}
\label{sec:data}

Twitter provides real-time access to tweets in the form of data streams over HTTP. Tweets are broadcast in the JSON format which encompasses meta-data in a hierarchical form. A tweet consists of not more that 140~unicode characters which allows for using any international script including east Asian languages. To each tweet, along with basic information, like date and time of the tweet, ID of the user etc., detailed information about the user is appended. User details contain fields such as the screen name, location (in a user-provided format that cannot be used for geo-tagging due to various reasons, see Sec.~\ref{sec:unknownloc}), date of registration as well as the number of tweets, friends and followers to the given date, etc. Tweets can be \textit{retweeted}, similarly to forwarding an e-mail. When a tweet is retweeted, the entry in the stream will contain the new tweet along with the original one, with detailed user information on both the original, and the retweeting user. Because any old tweet can be retweeted, streams might contain some old, historical tweets. When a tweet is a reply to another, the original tweet ID is provided. The Twitter system also parses out names of mentioned users from the text of tweets and resolves them to their numeric user IDs.

\subsection{Retweets and geo-tagged streams}

Users of Twitter broadcast tweets in the 10{,}000~s$^{-1}$ range which adds up to at least 2~Gbit~s$^{-1}$. While archiving and processing such a stream is feasible, a special subscription is required to access all tweets real-time. We instead collected data from the ``sprinkler'' stream which contains only a uniformly sampled 0.5\% of all tweets. The ``sprinkler'' stream contains about 50-60 tweets each second, or about 4-5~million tweets per day. The total collected amount of raw data is 10~GB/day, 2~GB/day when compressed.

A small fraction of tweets (not more than 1.5\%) is tagged with geographic information, usually in the form of real live GPS coordinates recorded by cell phones. A few users, such as weather forecast and earthquake watch services, etc., also publish tweets tagged with GPS information that are not necessarily the coordinates of the broadcasting station. As of Sept., 2013, Twitter allows downloading a filtered stream, which, when filtered to contain tweets from all around the world, provides about 50-60 geo-tagged tweets a second in real-time. The sampling factor of the geo-tagged stream is hard to be estimated as tweets the ``sprinkler'' stream barely contains any tweets that are also in the geo-tagged stream. While ``sprinkler'' contains many retweets, geo-tagged tweets are seldom retweets. This might be due to the fact that geo-tagged tweets are sent from cell phones which might not support retweeting as easily as the Twitter web site typically accessed from PCs without GPS receivers.

\subsection{Network data in the streams}
\label{sec:graphs}

There are three types of networks that can be readily extracted from the streams: 1) the ``retweet'' network, where edges among users are defined by one user retweeting a tweet of another; 2) the ``reply'' network, where edges come from replies; and 3) the ``mention'' network where edges are drawn if one user mentions another by their screen name. Networks also possess temporal information as each edge is stamped with the date and time of the tweet. Geographic information is often only available, if available at all, for one end of the edges. All three networks are directed but can be considered as non-directed, or one can require that edges point to both directions between two users. The database schema and pre-built indices we describe in Sec.~\ref{sec:schema} support all three (retweet, mention, reply) views of the networks, both directed and undirected.

\subsection{Follower graph discovery}
\label{sec:followergraph}

Twitter users can \textit{follow} (subscribe to) tweets of others. Following is an asymmetric connection between the two users, but the follower graph is considered the core information about the Twitter social network. For this reason, free access to the follower graph discovery interface is rather limited. Due to the value of this information, Twitter is constantly changing the means of gathering information about the follower network. We have been trying our best to keep up with these changes, so that we can collect as much of the follower graph as possible. By regularly querying the follower graph, we also expect to see some interesting temporal behavior in the network. Twitter has recently introduced significant changes to the graph discovery API which limits the number of nodes that can be visited in a day to about a mere 100{,}000. 

\begin{figure}
\begin{center}
\includegraphics[width=0.9\columnwidth]{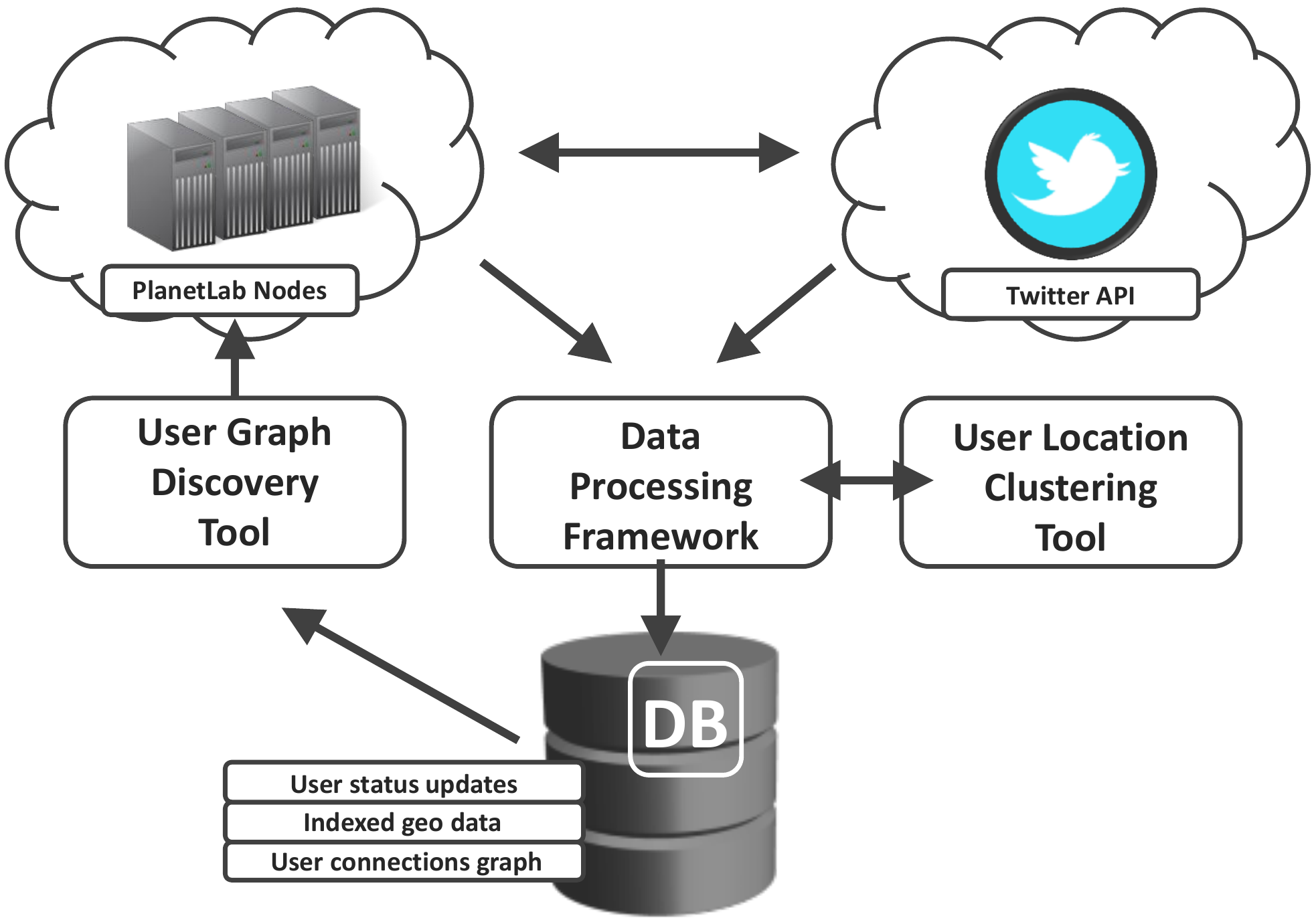} 
\end{center}
\vspace{-10pt}
\caption{Illustration of the data collection, data loader and graph discovery system, as it is build around a central database. The graph discovery tool can take the list of users from the database and pass the results to the data processing framework to ingest it into the database.}
\label{fig:loader}
\end{figure}

\section{Database}
\label{sec:database}

Our expertise with the product made Microsoft SQL Server~2012 an obvious choice as the platform for our data warehouse. SQL~Server has had built-in free-text search and geographic indexing support since version 2008 and the latest version also supports column-store indices which might significantly enhance the performance of graph analysis. Other database products were out of our interest due to various reasons. First of all, we built our solution over existing scientific tools \cite{DBLP:journals/corr/abs-cs-0701164, DBLP:dblp_conf/icws/OMullaneLNST05} that would have to be reimplemented from scratch for other platforms . Ad hoc query support, on the other hand, requires a SQL interface and a fast parallel engine to execute complex queries and scans, which ruled noSQL products and non-parallel databases out. The data warehouse was built on an eight-core commodity server with 16~GB of RAM and a specially tuned I/O subsystem with 40~TB of disk space providing maximum throughput of 1~GB~s$^{-1}$.

\subsection{Database schema}
\label{sec:schema}

The database schema storing Twitter data is relatively simple. The central fact table is ``tweet'' containing the short text messages. Tweets are organized into \textit{runs} to support storing the output from different Twitter streams, like the original ``sprinkler'' stream and the stream filtered to contain tweets with GPS coordinates. Beside the ``text'' field, the ``tweet'' table also contains a foreign key to the tweeting user and, if the tweet is a reply to a former tweet, the ID of the replied tweet.

A list of users is also maintained in a table called ``user''. The user table grows continuously as new users in the data streams are identified, and the table always contains the last known user profile status of each user. We also store historical information on user profiles in a table called ``user\_update''. As users eventually appear in the stream, we check for changes in the user profiles and create a record in the ``user\_update'' table for each change.

The text of each tweet is processed by Twitter and the streaming API provides us with the IDs of mentioned users and information on the retweeted tweet, if it is a retweet. We store these data in the ``tweet\_user\_mention'' and ``tweet\_retweet'' tables, respectively. These two tables, together with the ``tweet'' table, constitute the core of the social network data.

We provide views to access the three different networks of the data set. The view ``user\_mention'' lists the directed, time-stamped edges between the mentioned and the mentioning users. Views ``user\_reply'' and ``user\_retweet'' are defined similarly. Networks can be queried three different ways: 1) as time-stamped directed edges, 2) as non-directed, weighted edges, where any edge pointing to any direction of the original directed graph is considered as a connection between users, and 3) as mutual edges, where we require both directed edges to exist in the original directed graph. Indices on the corresponding tables are built to support fast querying of these views.

\subsection{Data loading procedure}

Twitter provides data as a ZIP compressed continuous stream of JSON documents that can be easily downloaded via HTTP. While downloading the stream is simple, converting it into a data warehouse is a fairly complex task. As a first step, the continuous stream is chopped into chunks containing about a day of data. The typical compressed size of a chunk containing raw JSON is 2.5-3~GB at the compression ratio of 5. It is important to mention that JSON documents can be compressed even at a ratio of 10 but compression/decompression of gzip streams is a single-threaded operation, consequently it can easily be the main bottleneck in the database loading pipeline.

Since we collect Twitter streams continuously, but data loading is a costly procedure in terms of both CPU and disk I/O, the database and the loading process had to be designed such a way that smaller amounts of new data can be efficiently merged with large amounts of old data. Data loading is done in batches, by chunks and in multiple steps. The loading process is implemented such a way that, shall an error occur, any chunk at any step can be continued.

The main steps of data loading are: 1) preparing bulk-load files for each batch, 2) bulk-inserting bulk-load files into a temporary database, 3) sorting daily data in the temporary database, 4) creating certain indices on the temporary data to support merging with the existing database, 5) merging data, 6) deleting temporary files and data tables, and 7) recreating indices on the large database.

Because reindexing the large database takes a significant time (a few hours), loading is usually done once in a month. The daily chunks are processed in parallel by a batch data loader program developed and optimized by us. Parallel processing of chunks is necessary to benefit from the multi-core system, but also to drive the high-speed I/O system efficiently. All file operations use large memory buffers (typically on the scale of tens of megabytes) to avoid random disk access. Preparing bulk-load files consists of parsing the JSON documents and writing data into a binary format that the database server can easily digest; one bulk-load file per data table per chunk is created. The bulk-load files are then ingested into a temporary database via simple bulk-inserts, without logging the transactions. Temporary tables are sorted by the same columns as the clustered indices of the table in the main database. This latter is absolutely necessary to make the merge process efficient.

Merging new data with old data is probably the hardest task, and requires lots of tricks and optimizations. Analyzing some of these tricks is rather instructive. As we mentioned above, new data has to be already sorted in the temporary database before the merging step. This typically can be done in a few seconds for a daily chunk. Merging with old data is different for different entities. Tweets are ordered by their monotonically increasing IDs but streams also contain historical tweets (in forms of retweets) that have smaller IDs. Consequently, these old tweets have to be inserted into the middle of the index structure. To avoid frequent B-tree page splits, which would reduce the performance significantly, the index fill factor of the tweet table is set to 80\%. This increases the size of the final database by 20\%, which also decreases net read-time I/O slightly, but speeds up the merging process by a factor of~5.

\begin{figure}
\begin{center}
\includegraphics[width=0.8\columnwidth]{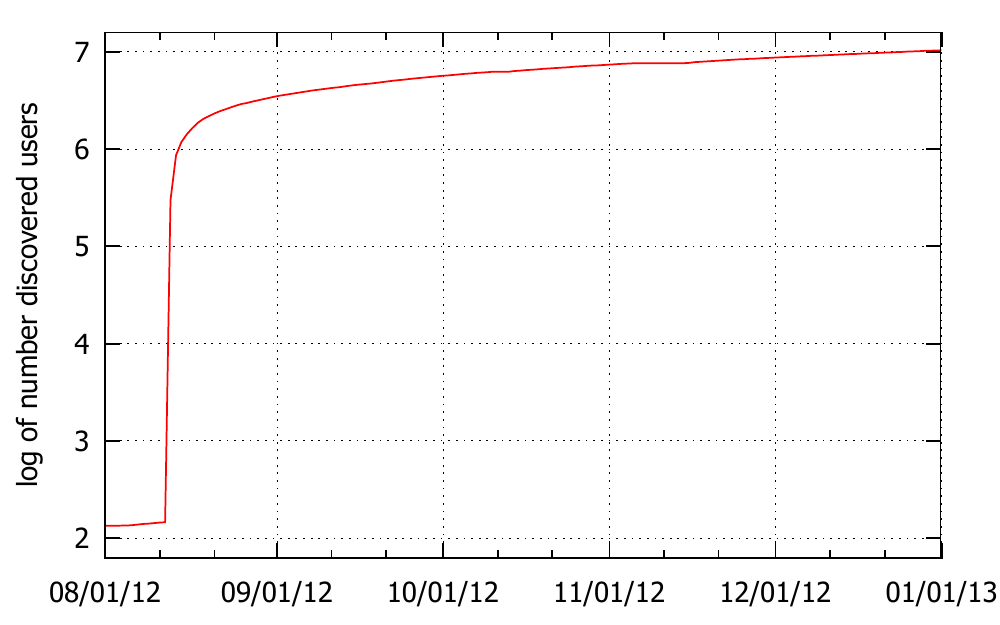} 
\end{center}
\vspace{-6pt}
\caption{Logarithm of the number of discovered users as a function of data collection date. The start of data collection time coincides with the very fast upturn of the curve. Some users are discovered from retweeted historical tweets, they make up the plateau before the start of data collection.}
\label{fig:userdiscovery}
\end{figure}

Page splits have even more significant effect in case of the ``user'' table, as user IDs can appear in the Twitter stream randomly. One would expect that the number of users yet unknown at a given time decreases as we discover more and more users over time, but this is not the case. In the first few days of data collection, user count grows according to $1-e^{-t/t_0}$ but the curve never flattens, rather it keeps increasing linearly, see Fig.~\ref{fig:userdiscovery}. It turned out that completely recreating the user table is faster than inserting new data into an already existing clustered index. To merge the ``user'' tables containing old and newly discovered users, we execute a join between them and write unique results into a new table with which the ``user'' table of the large database is subsequently replaced. In general, when the amount of new data is relatively small compared to old data, a lowered index fill-factor and using ordinary inserts is the way to execute a merge operation. When a significantly larger fraction of data is changed during merge, table recreation might be more beneficial.

Due to regular updates to the databases, file fragmentation is an issue that has to be addressed. We separated all major tables into various file groups, so that tweet, user and network edge information do not share the same data files. Also, the fill-factor of the clustered index on the tweet table is set to a 80\% which decreases file fragmentation when inserts are common. The user table is regularly recreated to keep the integrity of data pages.

\subsection{Indexing data}

For obvious performance reasons, during the merge process, non-clustered indices are disabled. Merge queries thus can rely solely on clustered indices. After merging in all chunks from a batch, non-clustered indices have to be rebuilt. To avoid the fragmentation of the database files due to continuous index rebuilds, indices are always created in dedicated file groups.

Indices on the ``tweet'' table are created for fast data retrieval by user ID and by geographic coordinates encoded as HTM IDs, see Sec.~\ref{sec:htm}. Primary keys of the tables storing network edges are ordered by the two endpoints of the edges. To allow query flexibility, inverse-ordered cover indices are created on these tables.


\subsection{Free text search}

The database is indexed for fast full-text search. We built our solution on the free-text indexing support of Microsoft SQL Server but implemented our own filter (in the form of a custom language filter) to efficiently eliminate noise words. By filtering out very short words (less than four characters), words containing numbers, words containing characters with high unicode values (asian script or characters used for decorating text) and words containing the same character consecutive more than twice, the size of the full-text index could be reduced by a factor of 10, when compared to the index size created with the default English language filter.

\subsection{Language identification}

Identifying the language of very short texts is challenging. According to our tests, simple ngram algorithms perform very badly on tweets. Windows Extended Linguistic Services, an API part of Windows since version~7, however, features a proprietary algorithm that performs much better than ngram. We identify the language of each tweet during the loading process and store it in the database. The language identification turned out to be robust enough to reproduce earlier work of others \cite{DBLP:dblp_journals/corr/abs-1212-5238}. Correlations between the structure of the social network, language use of individual users and the role of multi-lingual users will be an interesting area of research in the future.

\section{Handling geo-tagged data}
\label{sec:spatial}

Due to the wide-spread of smart phones, hundreds of millions of people carry GPS units with them that can also be connected to the Internet. Thousands of smart phone apps are available that can use GPS coordinates for various purposes: tag photos, collect sports data, provide localized advertisements and geo-tag social media contributions. Twitter supports geo-tagging tweets but, due to obvious reasons, disclosing GPS coordinates is a serious privacy issue, thus most users disable this feature. Nevertheless, we have identified more than ten million users who publish GPS coordinates regularly. Because geo-tagged tweets can be directly targeted by Twitter's streaming API, about 50\% of all tweets recorded by us are geo-tagged. The percentage of geo-tagged tweets of all tweets is somewhere around the 1.5\% level.

By processing geo-tagged data, spatial patterns in the behaviors of users can be identified. Such patterns can include daily commuting, living in two cities (college, home), going on vacation, etc. The large number of tweets, and the big scatter in GPS coordinates made it necessary to automatically identify clusters in coordinates on a per user basis.

\subsection{Spatial indexing of tweets}
\label{sec:htm}

To provide fast lookup of geo-tagged tweets by coordinates we index the data with the Hierarchical Triangular Mash (HTM) index \cite{raey, 2010PASP..122.1375B}. For every tweet with coordinates, we calculate the 20~level deep HTM ID which has roughly the resolution equivalent to an arc~second, or 25~meters. HTM was originally developed for indexing astronomical databases, hence its capabilities are determined by the needs of astronomers. While HTM is capable of indexing billions of data point, search regions \cite{DBLP:journals/corr/abs-cs-0701164} have to be relatively simple. On the other hand, the geographical indexing tools of SQL Server we also use to index certain kinds of data, see Sec.~\ref{sec:onthemap}, can handle very complex regions but indexing billions of rows with them is ineffective.

\subsection{Clustering coordinates}
\label{sec:clustering}

Users typically tweet from a few places where they spend most of their time \cite{Cho:2011:FMU:2020408.2020579}. These places are likely to be their homes, schools, workplaces, etc. To determine those few places where a user spends most of their time, we determine clusters in the GPS coordinates using the friend-of-friend (FoF) algorithm \cite{1982ApJ...257..423H}. The FoF algorithm is known from astronomy and widely-used to identify galaxy clusters. Two coordinates are considered to belong to the same cluster if their separation is less than 1~km. For each cluster, we determine the first two moments of the coordinate distribution. Before calculating the mean, to eliminate outliers, we trim data points until all points are inside a $3\sigma$ radius. We keep three clusters per user, the ones with the highest cardinalities. For each cluster, the average local hour of tweeting is calculated so that an estimate on the kind of the location can be made, for instance workplace or home.

\begin{figure}
\begin{center}
\includegraphics[width=0.8\columnwidth]{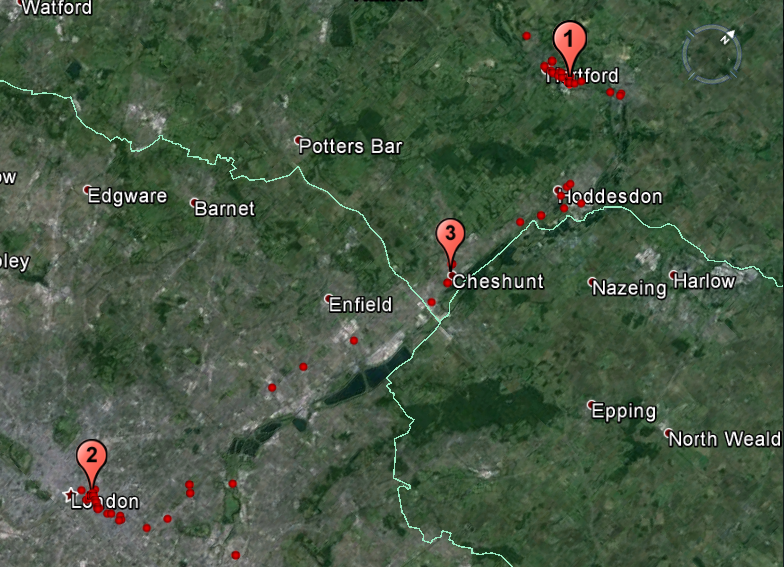} 
\end{center}
\caption{Clustered coordinates of tweets of a user. Only the three clusters with the highest cardinalities are kept. Map was created with Google~Earth.}
\label{fig:clusters}
\end{figure}

We prefer the friend-of-friend algorithm over other clustering techniques because, for example, FoF can identify any number of clusters while k-means expects the number of cluster as an input parameter. Also, the results of k-means can depend on the initially set cluster centers, while FoF always gives the same results. Furthermore, FoF can be parametrized by the value of separation, which allows for tuning cluster sizes and, the separation being the only parameter, the algorithm is rather simple. Clustering of tweet coordinates is done by an external tool that runs outside the database and outputs its result into a bulk-insert file for easy ingestion into the database. This technique turned out to be more efficient than trying to implement the FoF algorithm inside the database server. Fig.~\ref{fig:clusters} shows the results of the clustering of geo-tagged tweets of a user.

\subsection{Putting users on the map}
\label{sec:onthemap}

Since geographical embedding of social networks is in the center of our research, we focused on preparing our data warehouse for geospatial queries. As the database already contains about a billion tweets with GPS coordinates, and our performance requirements were high, precautions must have been taken when defining spatial indices. For this reason, we decided not to index the individual tweets, but rather index the GPS coordinate clusters we identified as described in Sec.~\ref{sec:clustering}. Clustering reduces the number of coordinates to be indexed from one billion to the order of 10~million. Once coordinate clusters are identified and average coordinates are determined, they get indexed using HTM and also by the built-in geographical index of SQL~Server.

\nopagebreak

When dealing with the geographical embedding of social networks, binning of the data by the coordinates is often necessary. Arbitrarily binning, however may not be the best way of performing an analysis, as bins may cut socially uniform areas, while other, socially diverse areas might be binned together. It is certainly not the best solution, but we decided to bin data according to administrative regions (countries, states, counties, etc.). For this purpose, we used the maps of administrative regions of the world from \url{gadm.org}, see Sec.~\ref{sec:gadm}.

Converting the maps to be used in the database is straightforward, finding the encompassing regions for tens of millions of GPS coordinates, however, is not. Looking up the geographic region containing a pair of coordinates is done by the server using spatial indices. First, the spatial index is used for coarse pre-filtering of the potential results, but however good the index implementation is, verifying the exact containment of a point inside a complex spherical polygon is a computationally intensive task. It turned out, that significant simplification of the region boundaries is necessary to make this task feasible, especially in the case of regions with seashores. Simplification of the region polygons, on the other hand, makes the boundaries ``fuzzy''. We may hope that erroneously classified coordinates will not effect the statistics of the regions as much as exact but arbitrarily chosen geographic binning.

\subsection{Importing the gadm.org database}
\label{sec:gadm}

To organize users according to political administrative regions, we downloaded the entire set of maps from \url{gadm.org}. The atlas contains all administrative regions of the world from the country level, sometimes down to the smallest villages. For our purposes, country, state and county levels are the most relevant. To load the map shapefiles into the database we used a freeware tool\footnote{Shape2SQL \url{http://sharpgis.net/page/Shape2SQL.aspx}} and relied on the geographical extensions of Microsoft~SQL~Server~2012.

\subsection{Users with unknown locations}
\label{sec:unknownloc}

Only about 1.5\% of Twitter users publish GPS coordinates. To determine the approximate location of users without GPS information statistically, one can rely on the geospatial information of the neighboring nodes of the social network \cite{Backstrom:2010:FMY:1772690.1772698}. Users can specify their locations in a textual form on the user profiles which, after careful filtering, can be also used to further refine geolocation estimates. Also, references to geographic locations within the tweets can be extracted and used. We will address this problem in details in a future paper.

\section{Data access and visualization}
\label{sec:access}

To simplify access to the database within our group, we adopted the web-based batch query system called CasJobs \cite{DBLP:dblp_conf/icws/OMullaneLNST05}, originally developed for the astronomical database SkyServer, and implemented several extensions to it that will eventually become part of CasJobs code base. To support our specific needs, we extended CasJobs with a schema browser to display the database structure and a new scriptable plotting tool. Due to legal reasons, access to the accumulated Twitter data is currently limited to research group members only.

\subsection{Adopting CasJobs}

CasJobs allows concurrent access to the data warehouse for multiple users via a batch system that can efficiently schedule long-running SQL queries. In CasJobs, users formulate their data reduction and analysis problems entirely in SQL, hence computations happen completely on the database server. This method helps avoid replicating the large datasets locally. The CasJobs infrastructure is hosted by a dedicated server containing a database for the SQL query batch service, the batch service itself, and sandbox databases of registered users called MyDBs. Query results automatically get stored in the MyDB, but users can also upload their own data to MyDB directly. Tables of the MyDB can be downloaded in various data formats.

\subsection{Visualization extensions to CasJobs}

The plotting tool developed by us is built around gnuplot\footnote{\url{http://www.gnuplot.info}} and allows users to directly plot results of SQL queries by using a slightly extended syntax of gnuplot scripts. This really makes the life of the data scientist easier because plots are generated on the database server, thus data do not have to be downloaded beforehand. Plotting scripts can also be saved which makes reproduction of plots much easier than with form-based graphic user interfaces. As the visualization module is based on gnuplot, almost all graphics formats supported by gnuplot are available, including postscript, JPEG, PNG and the ``canvas'' format which displays results as interactive HTML5 graphics.

\subsection{Map visualization in HTML5}

Along with the scriptable plotting tool, we also created a HTML5-based map display tool that enables us to interactively visualize and animate tens of thousands of tweets plotted over the map of a selected region of the world. The tool consists of a query panel where filtering of tweets can be specified, and a map view where the geographic and temporal distributions of the tweets are displayed. As with most browser-based client software, data is first downloaded from the data warehouse via a REST interface and displayed using client-side resources only. In our map visualisation tool, instead of bitmaps, map data is represented as polygons which allows for arbitrary projections and zooming on the client side, without transferring large amount of data from the server. A link to the map visualization tool is available on the website mentioned in the introduction.

\section{Sample queries}
\label{sec:queries}

To demonstrate the usefulness of the database, we consider a few queries calculating different statistical quantities of the streams and the networks. The first query simply determines the hourly distribution of tweets. The filter on the column ``run\_id'' restricts searches to the geo-tagged stream. The query obviously results in a table scan and executes in 2:01 with a sustained disc read of 950~MB~s$^{-1}$.

\begin{lstlisting}[language=sql]
SELECT DATEPART(hour, created_at), COUNT(*)
FROM tweet WHERE run_id = 1004
GROUP BY DATEPART(hour, created_at)
ORDER BY DATEPART(hour, created_at)
\end{lstlisting}
\vspace{-6pt}

The second query computes the histogram of the occurrence of the word ``network'', binned by days. This query benefits from the full-text index built on the ``text'' field of the ``tweet'' table. The query results in a series of index seeks and executes in 0:41.

\begin{lstlisting}[language=sql]
SELECT CAST(created_at AS date), COUNT(*)
FROM tweet WHERE run_id = 1004 AND
                     CONTAINS(text, 'network')
GROUP BY CAST(created_at AS date)
ORDER BY CAST(created_at AS date)
\end{lstlisting}
\vspace{-6pt}

The third query determines the degree distribution of outgoing edges of the ``mention'' graph. The simplicity of the query really shows the power of SQL for statistical purposes. This particular query translates into an index scan operation that can be readily aggregated. It completes in 5:57 for a graph with 765~million edges at a sustained read of 150~MB~s$^{-1}$.

\begin{lstlisting}[language=sql]
WITH degree_dist AS (
   SELECT user_id, COUNT(*) AS deg
   FROM tweet_user_mention WHERE run_id = 1004
   GROUP BY user_id )
SELECT deg, COUNT(*) FROM degree_dist
GROUP BY deg ORDER BY deg
\end{lstlisting}
\vspace{-6pt}

The fourth query calculates the number of tweets around New York City in a radius of 10~arc~minutes. This query uses the HTM index and requires a hint to achieve the best performance. Query execution results in an index scan and an aggregation and completes in mere 2 seconds.

\begin{lstlisting}[language=sql]
SELECT COUNT(*)
FROM dbo.fHtmCoverCircleEq(-74, 40.72, 10) htm
INNER JOIN tweet t
WITH (FORCESEEK(IX_tweet_htm(run_id, htm_id)))
ON t.htm_id BETWEEN 
               htm.HtmIDStart AND htm.HtmIDEnd
WHERE t.run_id = 1004
\end{lstlisting}

\nopagebreak

\section{Conclusions}
\label{sec:conclusions}

We have built a data warehouse of archival twitter data for data mining purposes. As an important feature, the database system supports efficient merging of new data with terabytes of existing data. We have shown that geo-tagged data can be successfully handled by the combination of the HTM indexing technique and the built-in geographic tools of Microsoft~SQL~Server.


\section*{Acknowledgment}
The authors thank the partial support of the European Union and the European Social Fund through project FuturICT.hu (grant no.: TAMOP-4.2.2.C-11/1/KONV-2012-0013), the OTKA 7779 and the NAP 2005/KCKHA005 grants. EITKIC\_12-1-2012-0001 project was partially supported by the Hungarian Government, managed by the National Development Agency, and financed by the Research and Technology Innovation Fund and the MAKOG Foundation.

\bibliographystyle{abbrv}
\bibliography{references}




\end{document}